\documentclass[prl, aps, 10pt, showpacs, superscriptaddress, twocolumn, floatfix]{revtex4-1}
\usepackage{graphicx}
\usepackage[usenames,dvipsnames]{color}

\newcommand{\HeIeenIpp}{$\mathrm{^{3}\vec{He}(\vec{\it e},{\it e'n}){\it pp}}$}
\newcommand{\HeIeepIpn}{$\mathrm{^{3}\vec{He}(\vec{\it e},{\it e'p}){\it pn}}$}
\newcommand{\DIeenIHrecoilalt}{$\mathrm{^{2}H(\vec{\it e},{\it e'\vec n})^{1}H}$}

\newcommand{\GEn}{\mbox{${G_{_{\!E}}^{n}}$}}
\newcommand{\GMn}{\mbox{${G_{_{\!M}}^{n}}$}}
\newcommand{\GE}{\mbox{${G_{_{\!E}}}$}}
\newcommand{\GM}{\mbox{${G_{_{\!M}}}$}}
\newcommand{\GEnGMn}{\GEn/\GMn}

\newcommand{\qstar}{\theta^*}
\newcommand{\fstar}{\phi^*}
\newcommand{\kinea}{2\sqrt{\tau(1+\tau)}\tan\frac{\theta_{\mathrm e}}{2}}
\newcommand{\kineb}{2\tau\tan\frac{\theta_{\mathrm e}}{2}\sqrt{1+\tau+(1+\tau)^2\tan^2\frac{\theta_{\mathrm e}}{2}}}

\newcommand{\kined}{\tau+2\tau(1+\tau)\tan^2\frac{\theta_{\mathrm e}}{2}}

\newcommand{\GeVsqr}{\mbox{GeV$^2$}}
\def\babar{\mbox{\sl B\hspace{-0.4em} {\scriptsize\sl A}\hspace{-0.4em} B\hspace{-0.4em} {\scriptsize\sl A\hspace{-0.1em}R}}}

\begin{document}

\title{Measurement of the neutron electric to magnetic form factor ratio at $Q^{2}=1.58\,\mathrm{GeV}^2$ using the reaction \HeIeenIpp}
\newcommand{\kph}{\affiliation{
    Institut f\"{u}r Kernphysik,
    Johannes Gutenberg-Universit\"{a}t Mainz, 
    D-55099 Mainz, Germany}}
\newcommand{\IfPMainz}{\affiliation{
    Institut f\"{u}r Physik,
    Johannes Gutenberg-Universit\"{a}t Mainz, 
    D-55099 Mainz, Germany}}
\newcommand{\zagreb}{\affiliation{
    Department of Physics, 
    University of Zagreb, 
    HR-10002 Zagreb, Croatia}}
\newcommand{\DoPBasel}{\affiliation{
    Department of Physics, 
    University of Basel,
    CH-4056 Basel, Switzerland}}
\newcommand{\tuebingen}{\affiliation{
    Physikalisches Institut, 
    Universit\"at T\"ubingen,
    D-72076 T\"ubingen, Germany}}
\newcommand{\stefan}{\affiliation{
    Jo\v{z}ef Stefan Institute, 
    SI-1000 Ljubljana, Slovenia}}
\newcommand{\ljubljana}{\affiliation{
    Department of Physics, 
    University of Ljubljana, 
    SI-1000 Ljubljana, Slovenia}}  
\newcommand{\fiu}{\affiliation{
    Department of Physics,
    Florida International University,
    Miami FL 33199, U.S.A.}}
\newcommand{\clermont}{\affiliation{
    Clermont Universit\'e, Universit\'e Blaise Pascal, 
    F-63000 Clermont-Ferrand, France}}
\newcommand{\affpsi}{\affiliation{
    Paul Scherrer Institut,
    CH-5234 Villigen, Switzerland}}
\newcommand{\Lyon}{\affiliation{
    Universit{\'e} Lyon 1,
    F-69622 Villeurbanne, France}}

\author{B.\,S.~Schlimme} \email[]{schlimme@kph.uni-mainz.de}\kph\homepage{http://www.kph.uni-mainz.de}
\author{P.~Achenbach}\kph
\author{C.\,A.~Ayerbe~Gayoso}\kph
\author{J.\,C.~Bernauer}\kph
\author{R.~B\"ohm}\kph
\author{D.~Bosnar}\zagreb
\author{Th.~Challand}\DoPBasel
\author{M.\,O.~Distler}\kph
\author{L.~Doria}\kph
\author{F.~Fellenberger}\kph
\author{H.~Fonvieille}\clermont
\author{M.~{G\'omez~Rodr\'iguez}}\kph
\author{P.~Grabmayr}\tuebingen
\author{T.~Hehl}\tuebingen
\author{W.~Heil}\IfPMainz
\author{D.~Kiselev}\affpsi
\author{J.~Krimmer}\IfPMainz\Lyon
\author{M.~Makek}\zagreb
\author{H.~Merkel}\kph
\author{D.\,G.~Middleton}\tuebingen
\author{U.~M\"uller}\kph
\author{L.~Nungesser}\kph
\author{B.\,A.~Ott}\tuebingen
\author{J.~Pochodzalla}\kph
\author{M.~Potokar}\stefan
\author{S.~{S\'anchez Majos}}\kph
\author{M.\,M.~Sargsian}\fiu
\author{I.~Sick}\DoPBasel
\author{S.~\v{S}irca}\stefan\ljubljana
\author{M.~Weinriefer}\kph
\author{M.~Wendel}\kph
\author{C.\,J.~Yoon}\kph

\date{\today}

\begin{abstract}
A measurement of beam helicity asymmetries in the reaction \HeIeenIpp \ has been performed 
at the Mainz Microtron 
in quasielastic kinematics 
in order to determine the electric to magnetic form factor ratio of the neutron, \GEnGMn, at a four momentum transfer $Q^2=1.58\,\mathrm{GeV}^2$.
Longitudinally polarized electrons were scattered on a highly polarized $^3$He gas target. The scattered electrons were detected with a high-resolution magnetic spectrometer, and the ejected neutrons with a dedicated neutron detector composed of scintillator bars.
To reduce systematic errors data were taken for four different target polarization orientations allowing the determination of $\GEnGMn$ from a double ratio.
We find $\mu_{\it n}\GEnGMn=0.250 \pm 0.058\mathrm{(stat.)} \pm 0.017\mathrm{(sys.)}$.
\end{abstract}


\maketitle

\paragraph{Introduction---\hspace{-3mm}}

The electromagnetic structure of the nucleons can be probed systematically by electron scattering experiments.
The $Q^2$ dependence of the electric and magnetic Sachs form factors (FF) $\GE$ and $\GM$ of the proton and the neutron reflect the distributions of charge and magnetization inside the nucleons, respectively.
Precise measurements of the form factors over a wide $Q^2$ range are well suited for testing non-perturbative QCD and structure models and therefore are essential for a quantitative understanding of nucleon structure.

Elastic form factors of the free neutron can be deduced from scattering experiments on light nuclei in quasifree kinematics.
The magnetic neutron form factor $\GMn$ has been measured in unpolarized scattering reactions up to moderately high $Q^2$ values with small errors (e.g. \cite{Lachniet} and references therein).
Measurements of $\GEn$ are very difficult in unpolarized reactions since $\GEn$ is small due to the vanishing net charge of the neutron, thus contributions of $\GEn$ to unpolarized cross sections are small.
Sensitivity to $\GEn$ can be tremendously enhanced in double polarization experiments where polarized electrons scatter quasielastically on deuterons or $^3$He, and either the target is polarized or the polarization of the ejected neutrons is determined \cite{DonnellyRaskin}. 
The FF ratio $\GEnGMn$ has been studied in several such experiments \cite{Riordan,RohePRL,*BermuthPhysLett,Madey,*Plaster,Glazier,Herberg,*Ostrick,Eden,Geis,Warren,Zhu,Passchier,Meyerhoff,Becker,*BeckerGolak,Thompson,JonesWoodward} from which $\GEn$ could be determined. 
For low $Q^2$ the momentum transfer dependence of $\GEn$ can be well described with a parametrization which was originally introduced in \cite{galster} to represent $\GEn$ found in unpolarized {\it ed} scattering.
However, in \cite{Madey,*Plaster}
a considerable deviation was found in analysis of asymmetries in the reaction \DIeenIHrecoilalt \ at a previously unreached high $Q^2=1.45\,$\GeVsqr.
Two experiments have been performed in this $Q^2$ region to investigate $\GEnGMn$ using a different reaction, \HeIeenIpp, 
which is sensitive to different systematic uncertainties: One at Jefferson Lab (JLAB) and the one at the Mainz Microtron (MAMI) discussed here. While at JLAB data were taken up to $Q^2=3.4\,$\GeVsqr \ \cite{Riordan}, the main focus of the Mainz experiment was a suppression of systematic uncertainties 
by exploiting ratios of asymmetries measured for different kinematics. 
The resulting systematic error is smaller by almost a factor of 2 compared to \cite{Riordan}, whereas the statistical uncertainty is larger due to different kinematics, smaller luminosity and detector acceptances.

\paragraph{Measurement technique---\hspace{-3mm}} \label{intro}

Due to its special spin structure polarized $^3$He can be used as an effective polarized neutron target with a high relative neutron polarization, while the mean proton polarization is small \cite{Blankleider,Scopetta2007}.

Beam helicity asymmetries $A=\frac{N^+-N^-}{N^++N^-}$ of the reaction \HeIeenIpp\ have been measured in quasielastic kinematics with longitudinally polarized electrons. 
$N^+$ and $N^-$ denote the normalized counts at positive and negative electron helicities, respectively.

\begin{figure}
  \begin{center}
    \includegraphics[angle=0, width=\columnwidth]{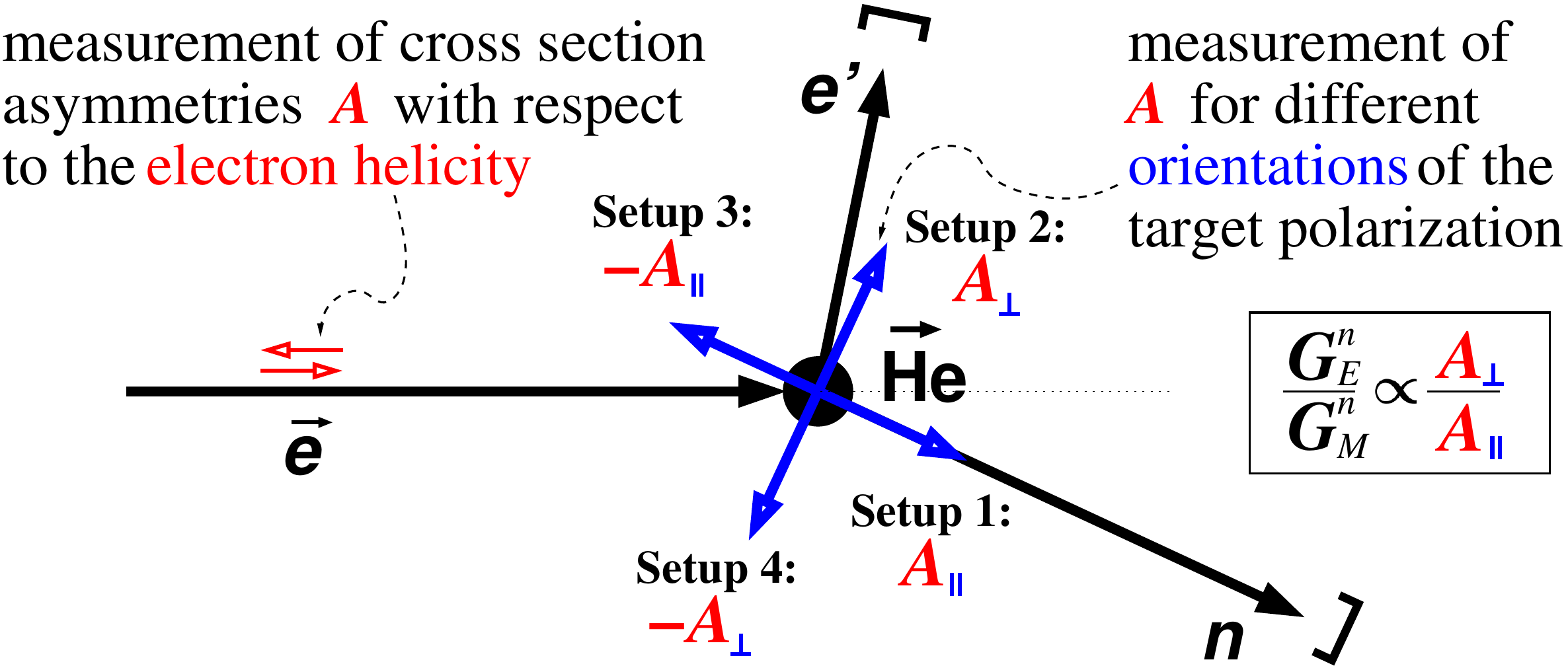}
    \caption{(Color online) Asymmetries of the reaction \HeIeenIpp \ have been measured for four different target polarization orientations. The asymmetries obtained in Setup 2 and Setup 4, that correspond to a polarization direction in the scattering plane and perpendicular to the (mean) momentum transfer ($\qstar=90^\circ; \fstar=0^\circ$ and $180^\circ$, respectively), are sensitive on $\GEnGMn$. Asymmetries measured in Setup 1 and Setup 3 ($\qstar=0^\circ$ and $180^\circ$) are used for normalization.\label{fig:MT}}
  \end{center}
\end{figure}

Data analysis was based on the assumption that scattering took place on a free neutron at rest. Nuclear binding effects have been studied subsequently.
In this case $A$ is given in the one-photon exchange approximation from \cite{DonnellyRaskin}
\begin{eqnarray}
  A(\qstar, \fstar) & = & -\frac{ag\cdot \sin\qstar\cos\fstar + b\cdot \cos\qstar}{g^2+d} \cdot P_{\it e}P_{\it t}V\label{eq:A_freeNeutron}, \\
  g & = & \GEnGMn, \nonumber\\
  a & = & \kinea,  \nonumber\\
  b & = & \kineb,  \nonumber\\
  d & = & \kined. \nonumber
\end{eqnarray}
$P_{\it e}$ and $P_{\it t}$ are the electron and the target polarizations.
$\theta_{\mathrm e}$ is the electron laboratory scattering angle, $\tau=Q^2/(4M^2)$ with the nucleon mass $M$. 
$\qstar$ and $\fstar$ denote the neutron spin orientation with respect to the momentum transfer $\vec q$ and the scattering plane, respectively.
The quantities $a$, $b$ and $d$ are given by the electron kinematics and are of the order $1$, i.e. $\gg |g|$. The factor $V$ accounts for a possible dilution due to contributions with vanishing asymmetry.
High sensitivity to the FF ratio $g$ is obtained for asymmetries $A_\perp$ measured at a 
neutron spin orientation perpendicular to the momentum transfer in the scattering plane ($\qstar=90^\circ, \fstar=0^\circ$, see Fig.\ \ref{fig:MT}). 

Furthermore, asymmetries $A_\parallel$ with a parallel orientation can be used for normalization leading to 
$g=b/a \cdot \frac{\left(P_eP_tV\right)_\parallel}{\left(P_eP_tV\right)_\perp}A_\perp/A_\parallel$. 
In this case there is no direct need for absolute polarization determinations.
Note also that only the helium polarization can be measured directly. To extract the absolute neutron polarization, the relative neutron to helium polarization must be known. In the asymmetry ratio, this relative polarization factor drops out, as well as for instance a dilution factor of the neutron polarization due to any admixture of nitrogen to the target gas needed for technical reasons. Finally it should be highlighted that helicity independent experimental background also cancels in the ratio within statistical fluctuations, i.e. $V_\perp=V_\parallel$ since $V$ does not depend on the orientation of the target spin.

\paragraph{Experiment---\hspace{-3mm}}

The experiment was carried out at the spectrometer setup of the A1 Collaboration (see \cite{Blomqvist98} for a detailed description) at MAMI \cite{Herminghaus:1976mt}. 
MAMI-C \cite{Kaiser:2008} provided $1.508\,$GeV longitudinally polarized electrons \cite{Aule97} at a beam current of $10\,\mu$A. 
The beam polarization was determined twice per day using a M\o ller polarimeter.
Only minor fluctuations around a mean polarization of $76.3\,\%$ were found.
The beam helicity was flipped quasi-randomly once per second.
The electron beam impinged on a polarized $5\,$bar $^3$He gas target, resulting in a typical luminosity of $4\cdot 10^{34}\,\mathrm{cm^{-2}\,s^{-1}}$.
Due to the sophisticated setup, it was possible to turn a magnetic guiding field at the target in any direction while keeping the magnetic gradient sufficiently low despite of the magnetic stray field of the spectrometer.
Therefore the target polarization could be oriented in any desired direction. 
Data were taken for the four orientations depicted in Fig.\ \ref{fig:MT}.
The target was polarized at the nearby Mainz Institute of Physics 
with an initial polarization of up to $72\,\%$. The polarized gas was stored in cesium coated quartz glass cells which were transported to the target area in a holding field. The target cells were exchanged twice per day. With a relaxation time of around $30$ to $40$ hours under beam conditions, this resulted in a mean polarization of $55.6\,\%$. Details on the target setup, polarization measurements and performance during the beam time can be found in \cite{Krimmer:2009zz}.

The scattered electrons were detected with a high resolution spectrometer with a $28\,$msr solid angle and a $5\,$cm target length acceptance.
Vertical drift chambers were used for tracking, scintillator detectors for trigger and timing purposes, and a threshold-gas-\v{C}erenkov detector for discrimination between electrons and pions in off-line analysis.
This setup has a relative momentum resolution of $\approx 10^{-4}$ and an angular resolution at the target of $\approx 3\,$mrad \cite{Blomqvist98}.

For coincident detection of the recoil nucleons a plastic scintillator array was used. It consisted of six layers of five detector bars of size $50 \times 10 \times 10\,\mathrm{cm^3}$. The bars were equipped with two photomultiplier tubes (PMTs) providing signal height and timing information. Copper plates with a thickness of $2\,$cm were inserted between the layers to increase the neutron detection efficiency. Two layers of $1\,$cm thick veto bars were installed in front of the detector bars for rejection of charged particles in off-line analysis. These each had one PMT attached. 
The neutron detector was shielded with $5\,$cm of boron-treated polyethylene and $10\,$cm of lead except for an entrance window pointing to the target that was shielded with $1\,\mathrm{cm}$ of lead to keep the charge conversion probability small for protons coming from the target.

The event trigger was formed by an $80\,$ns wide coincidence between the signals of the spectrometer scintillator paddles and the neutron detector bars, the veto bars of the neutron detector were not included.
Apart from the measurements on a $^3$He-target, data were taken on a hydrogen target to estimate the proton rejection efficiency. Background coming from electron scattering on the target cell entrance foils was studied in empty cell measurements and was found to be negligible.

The central kinematics of this experiment is summarized in Tab.\ \ref{tab:centKine}.
\begin{table}[h!tb]
  \caption{Central kinematics of the experiment. $Q^2=\vec q^2-\omega^2$ the negative of the squared four-momentum transfer, $\omega$ the energy transfer, $\vec q$ the three-momentum transfer, $k_0$ the beam energy, $k_0'$ the energy of the scattered electron, $\theta_{\mathrm e}$ the electron scattering angle and $\theta_{\mathrm n}$ the angle of the ejected neutron.}\label{tab:centKine}
  \begin{center}
    \begin{tabular}{c|c|c|c|c|c|c}
      $Q^2$ & $\omega$ & $\left|\vec q\right|$ & $k_0$ & $k_0'$ & $\theta_{\mathrm e}$ & $\theta_{\mathrm n}$\\ 
      $[\mathrm{GeV^2}]$ & $[\mathrm{GeV}]$ & $[\mathrm{GeV}]$ & $[\mathrm{GeV}]$ & $[\mathrm{GeV}]$ & $[^\circ]$ & $[^\circ]$\\ \hline
      $1.58$ & $0.855$ & $1.52$ & $1.508$ & $0.653$ & $78.6$ & $24.9$ 
    \end{tabular}
  \end{center}
\end{table}

\paragraph{Data analysis---\hspace{-3mm}}

The well known properties of the spectrometer optics were used to reconstruct the electron momentum, its direction and the reaction vertex. The momenta of the incoming (given by the beam energy) and outgoing electrons were corrected for energy loss in material along their track and due to bremsstrahlung by the calculated most probable value.
The timing and amplitude signals of the neutron detector bars were used to reconstruct the direction of the recoiling particles.
After electron identification, that was mainly accomplished through the evaluation of the \v{C}erenkov detector signals, the coincidence time spectrum between electron and neutron arms was very clean. After timing calibration a peak width of $2.1\,$ns FWHM was achieved. A $5\,$ns wide cut on the coincidence time was used, the fraction of accidental coincidences was estimated by a side band analysis to be $0.7\,\%$.

The invariant mass $W$ of the virtual photon and the target nucleon was calculated from electron kinematics under the assumption of a free target nucleon at rest as $W=\sqrt{(\omega + M)^2-\vec q^2}$. It was used for a suppression of inelastic events via a cut on $W<1030\,$MeV. To keep corrections due to nuclear binding effects small, 
it is additionally desirable to select data with low missing momentum $\vec p_{\mathrm{miss}}=\vec q - \vec p_{\it N}$ where $\vec p_{\it N}$ is the nucleon momentum. The neutron detector was too close to the target ($2.1\,$m) for a reasonable determination of $\left |\vec p_{\it N}\right |$ via time-of-flight. Due to this a cut on $q^\perp=q\cdot \tan\alpha<150\,$MeV with the reconstructed angle $\alpha$ between $\vec p_{\it N}$ and $\vec q$, which is strongly correlated to the missing perpendicular momentum, was applied, yielding a resolution of $\approx 20\,$MeV.

\begin{figure}
  \begin{center}
    \includegraphics[angle=270, width=\columnwidth]{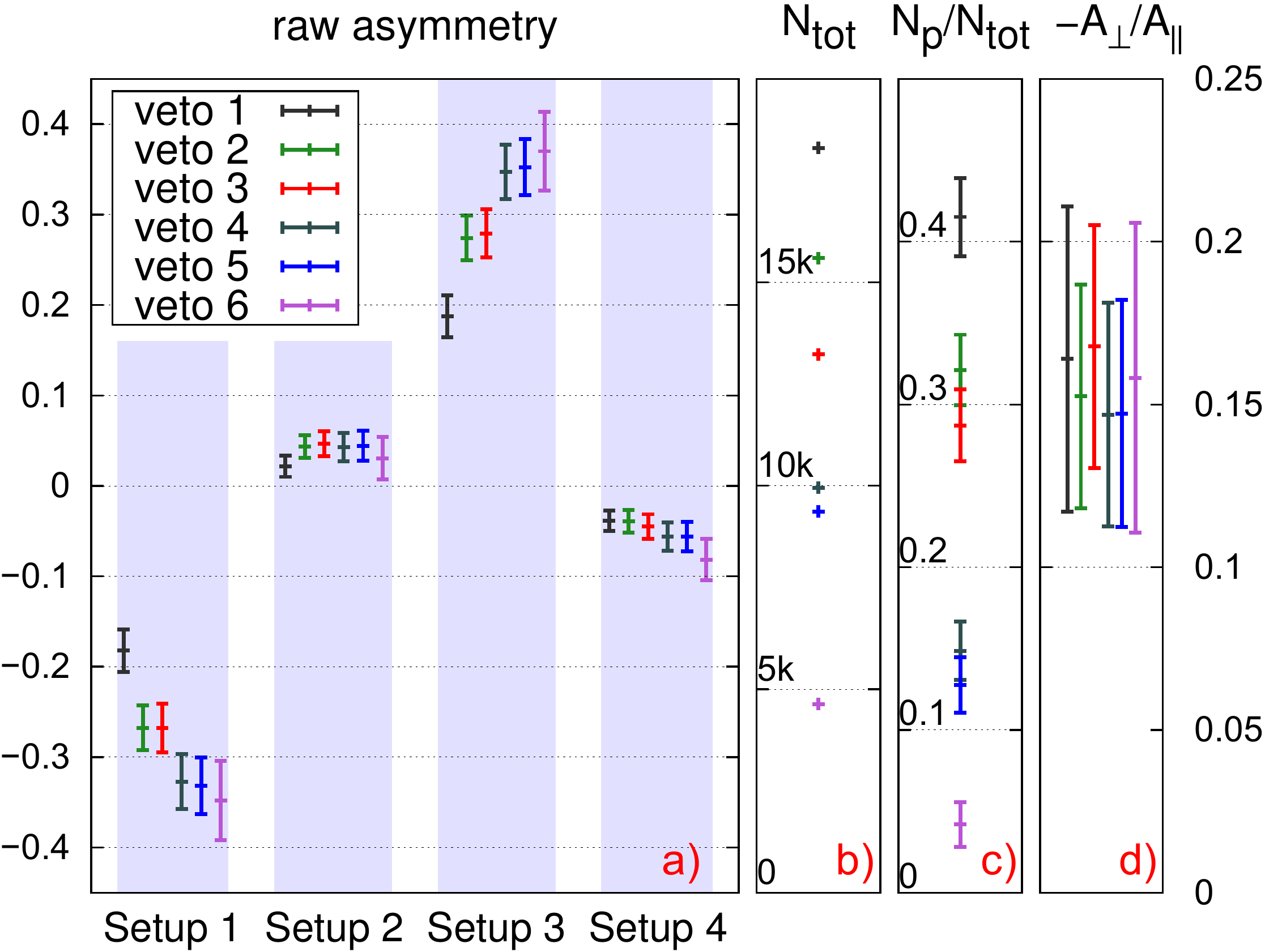}
    \caption{(Color online) {\bf a)} Experimental raw asymmetries obtained in the four setups for different proton rejection criteria without any corrections. 
For instance {\it veto 1} denotes the asymmetries obtained when only the timing signals of the neutron detector veto counters were used for proton rejection. For analysis {\it veto 5} was used which included an additional cut on the reconstructed energy deposited in the detector bars of the first layer.
{\bf b)} Total number of accepted events. {\bf c)} Relative proton contamination estimated by analysis of scattering data on a hydrogen target. {\bf d)} Ratio of $A_\perp$ (statistically weighted mean value of the two asymmetries measured in Setup 2 and Setup 4) and $A_\parallel$, which is essentially used to extract $g$.\label{fig:differ}}
  \end{center}
\end{figure}

For discrimination between protons and neutrons the veto counters of the neutron detector were used. However a high level of electromagnetic background from the target affected their signal quality.
A proton rejection based on the signals of these bars alone was found to be rather inefficient.
In order to reduce the proton contribution in a more efficient way
the first layer of the detector bars was additionally used for charged particle rejection. 
Many different proton rejection criteria were tested, the resulting raw asymmetries for some of those (representative) are shown in Fig.\ \ref{fig:differ}. 
The robustness of this measurement method concerning proton dilution becomes apparent:
Since the mean polarization of the protons is small, their (measurable) cross section asymmetry with respect to the beam helicity is small. 
The proton contribution can lead to a significant dilution of the measured asymmetries. But since the asymmetries measured in the different setups are diluted by almost the same factor, the residual net effect to the asymmetry ratio is comparatively small.
For further analysis, a simple criterion was chosen which was not too restrictive but still delivered a good proton rejection: In addition to a cut on the timing of the veto bars (which was viable if present), a cut on the reconstructed energy deposition in the detector bars of the first layer was applied below the proton peak which was clearly visible in the accumulated spectra.

In the following an event-by-event analysis was performed based on a maximum likelihood (ML) fit. It was used to find the values for the two parameters $g$ and $V$, common for all setups. The factor $V$ includes the effective neutron polarization inside the $^3$He.
$g$ was assumed to be constant over the $Q^2$ range ($\Delta Q^2=0.08\,$\GeVsqr \ RMS).
Starting from a ML-fit appropriate to take into account asymmetries following Eq.~\ref{eq:A_freeNeutron}, effects that can lead to a deviation of the individual asymmetries were subsequently included.
The reconstructed kinematic quantities including $\qstar$ and $\fstar$ were considered explicitly for each event as well as current electron and target polarizations. 
Helicity dependent beam current fluctuations, which had been monitored, and accidental coincidence rates were integrated, as well as the helicity dependent proton contribution. 
The relative proton contamination due to misidentification and charge exchange reactions was estimated to $(12.8\pm1.7)\%$ by an auxiliary analysis of scattering data on a hydrogen target using the same cuts as for the $^3$He data.
The helicity dependence of that contamination was deduced for the individual target polarization settings from analysis of reconstructed events in the reaction \HeIeepIpn. There, the mean polarization of the protons relative to the $^3$He polarization was found with a similar ML-fit to be $P_{\it p}/P_{\mathrm{^3{He}}}=-0.038\pm0.027$ when the proton FF parametrization of \cite{Kelly} was used. Taking the proton contribution into account led to a slight reduction of the extracted value for $g$ by $1.6\,\%$.

The final result of the fit was $g=-0.136\pm0.031$. The statistical error of $g$ was determined via the inverse of the covariance matrix.

The influence of effects like Fermi motion, detector resolutions and energy loss of the particles in material or due to bremsstrahlung was investigated by using a Monte Carlo simulation in the framework of the plane wave impulse approximation (PWIA).
The acceptances of the detectors and the applied kinematic cuts were all taken into account.
The standard dipole parametrization $G_{\mathrm{D}}(Q^2) = \left(1+\frac{Q^2}{0.71\,\mathrm{GeV}^2}\right)^{-2}$ and a Galster like parametrization \cite{FriedrichWalcher} matching our extracted $\GEn$ were assumed for $\GMn/\mu_{\it n}$ and $\GEn$, respectively.
Pseudodata were generated and were fed to a ML-fit similar to the one used for the data analysis to study the individual effects on the fitting procedure used for data analysis. 
An overall correction factor leading to a reduction of the extracted value for $g$ from the ML-fit by $3.6\,\%$ \ was found.

The influence of final state interactions (FSI) was studied by performing calculations including 
spin dependent FSI within the generalized eikonal approximation (GEA) \cite{GEA}.
These calculations use the SAID parametrization \cite{SAID} of the spin dependent nucleon-nucleon scattering amplitudes, which is also used to describe the charge-interchange proton-neutron rescattering. The FSI between two outgoing slow protons is calculated in the pair-distortion approximation based on the nonrelativistic quantum mechanical scattering approximation \cite{GEA}.
The calculation from \cite{3NBoundStates} employing the AV18 nucleon-nucleon potential \cite{AV18} was used for construction of the $^3$He ground-state wave function.
After integration over a set of kinematics covering our experimental acceptance, 
the calculated asymmetries were compared with the results obtained in PWIA.
A negligible deviation of the asymmetries both for perpendicular and parallel target spin alignments was found leading to a negligible total effect on the determination of $g$.

For an estimate of the influence of inelastic events a simulation 
employing the MAID model \cite{MAID99,*MAID07} for single pion electroproduction within the PWIA
was performed. The yields for events containing a neutron in the final state were compared to the simulation results for quasielastic events. The inelastic fraction was estimated to be 1.6$\%$ relative to the number of quasielastic scattering events. Under consideration of the mean electron and nucleon polarizations but without any assumption on the physical asymmetries associated with these events, a resulting uncertainty of $4.8\,\%$ is conservatively estimated.

\paragraph{Result and Discussion---\hspace{-3mm}}

As a final result $\mu_{\it n}\GEnGMn=0.250 \pm 0.058\mathrm{(stat.)} \pm 0.017\mathrm{(sys.)}$ is found.
The statistical uncertainty is given by the error estimation of the ML-fitting procedure, the systematic uncertainty is the sum of the individual uncertainties, listed in Tab.\ \ref{tab:syserrors}, added in quadrature. 
Using the value for $\GMn$ from the standard dipole parametrization at $Q^2=1.58\,$\GeVsqr\ with an uncertainty of $2\,\%$ reflecting the level of agreement with data from \cite{Lachniet} yields
$\GEn=0.0240\pm0.0056\mathrm{(stat.)}\pm0.0017\mathrm{(sys.)}$ for the electric form factor of the neutron.

\begin{table}[h!]
  \caption{Contributions to the systematic uncertainty of $g=\GEnGMn$ from individual sources (relative to the $g$-value).}\label{tab:syserrors}
  \begin{ruledtabular}
  \begin{tabular}{l|l}
    source & $(\Delta g)^{\mathrm{sys}}/g$\\ \hline
    residual proton contribution & $1.2\,\%$\\
    Fermi motion, detector resolutions, energy loss \quad & $1.2\,\%$\\
    accuracy GEA calculations & $2.0\,\%$\\
    residual pion production event contribution & $4.8\,\%$\\
    accuracy target polarization alignment & $3.8\,\%$\\
    electron polarization & $0.2\,\%$\\
    target polarization & $0.8\,\%$\\ 
    other & $1.1\,\%$\\
  \end{tabular}
\end{ruledtabular}
\end{table}

\begin{figure}
  \begin{center}
    \includegraphics[angle=0, width=\columnwidth]{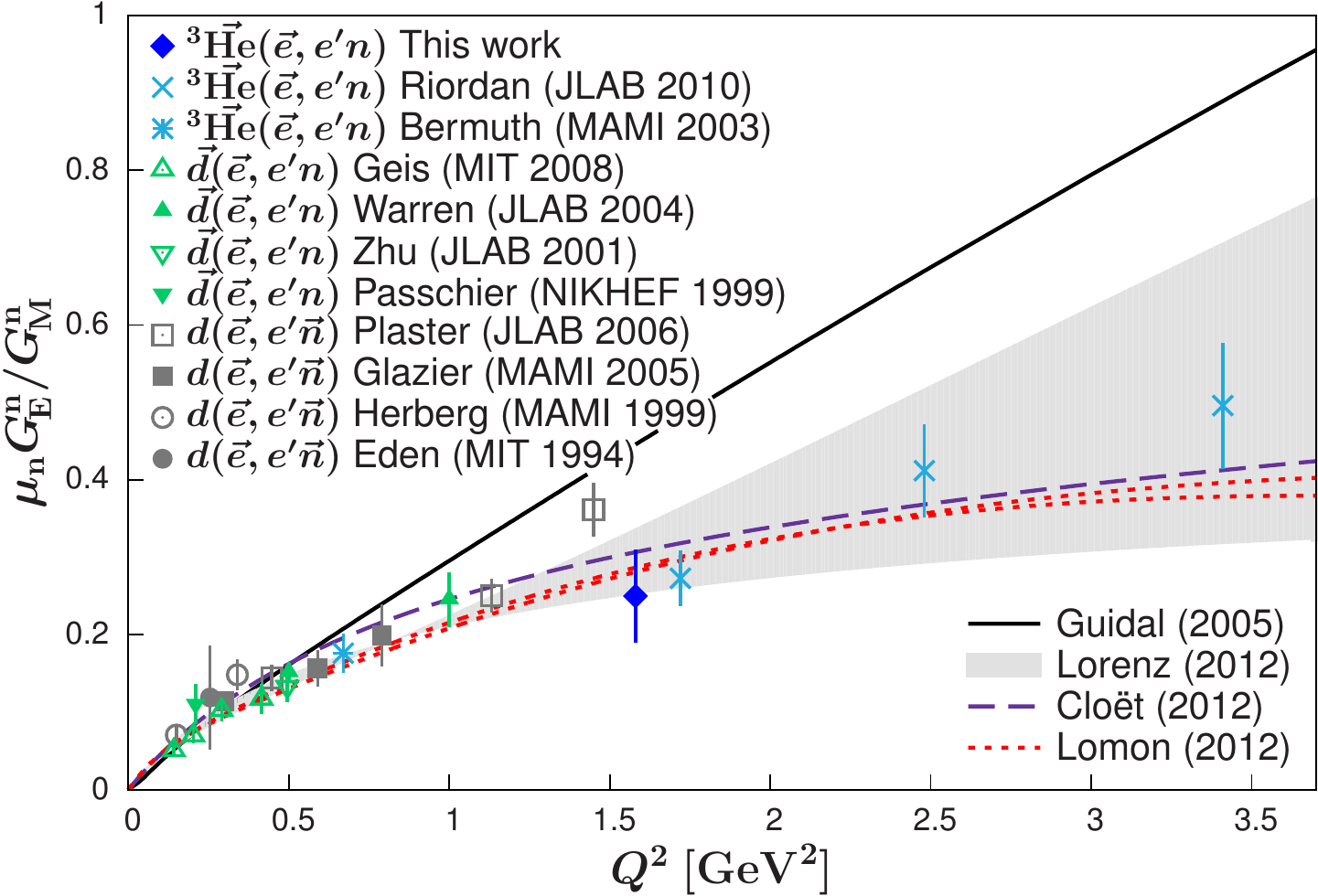}%
    \caption{(Color online) The result for $\mu_{\it n}$\GEnGMn \ of this experiment along with results from other double polarization experiments \cite{Riordan,RohePRL,*BermuthPhysLett,Madey,*Plaster,Glazier,Herberg,*Ostrick,Eden,Geis,Warren,Zhu,Passchier}. The uncertainties shown are the statistical and the systematic errors added in quadrature.
Also shown are the results of recent calculations based on general parton distributions \cite{gGPD} (solid line), dispersion analysis \cite{Lorenz} (gray band), a quark-diquark model with a pion cloud \cite{gquarkdiquark} (dashed line) and the extended Lomon-Gari-Kr\"umpelmann model of nucleon electromagnetic FF \cite{gVMD} (dotted lines, for two different parametrizations of resonance widths, timelike proton form factor data from \babar \ included).\label{fig:g}}
  \end{center}
\end{figure}

The present result for $\mu_{\it n}$\GEnGMn \ is shown in Fig.\ \ref{fig:g} together with a selection of published results from other double polarization experiments.
The precise measurement of \cite{Madey,*Plaster} as well as several calculations (for instance \cite{gGPD}, which is also shown in the figure) indicated a steep slope of the form factor ratio at intermediate $Q^2$. 
In contrast to this, the results of \cite{Riordan} do not support this trend. 
Our experiment used the same reaction as \cite{Riordan}, but in different kinematics and with a different setup, that allowed us to determine the form factor ratio from a double ratio for a reduction of systematic errors. Our result 
is in good agreement with the result of \cite{Riordan} which was measured at a similar $Q^2$, and therefore confirms that the FF ratio slope is shallower.

\begin{acknowledgments}
We gratefully acknowledge the support from the technical staff at the Mainz Microtron and thank the accelerator group for the excellent beam quality. 

This work was supported by the Deutsche Forschungsgemeinschaft with the Collaborative Research Center 443 and by the Europ\"aisches Graduiertenkolleg ``Hadronen im Vakuum, Kernen und Sternen'' Basel-Graz-T\"ubingen.
\end{acknowledgments}

\bibliographystyle{apsrev4-1} 
\bibliography{g}
\end{document}